\documentstyle[12pt,epsfig]{article}
\def\thebib#1{
 \list
 {\arabic{enumi}}{\settowidth\labelwidth{[#1]}\leftmargin\labelwidth
 \advance
 \leftmargin\labelsep
 \setlength{\parsep}{0mm}%
 \setlength{\itemsep}{0mm}%
\usecounter{enumi}}
 \def\newblock{\hskip .11em plus .33em minus .07em}
 \sloppy\clubpenalty4000\widowpenalty4000
 \sfcode`\.=1000\relax}

\renewcommand{\thefootnote}{\fnsymbol{footnote}}

\def \invisible{\mbox{$\rule{0mm}{1mm}$}}
\def \mathbox(#1){\invisible\ifmmode{{#1}}\else{\mbox{${#1}$}}\fi}
\def \mbf(#1){\mbox{\boldmath{$#1$}}}

\def\mbfp{\mbf(p)}

\def\calO{{\cal O}}
\def\calA{{\cal A}}

\def\III{{\rm I\thinspace I\thinspace I}}

\def\etal{{\it et al.}}

\def\piii{p_{\rm III}}

\def \mbfp{\mbf(p)}

\newcommand{\bra}{\langle}
\newcommand{\half}{\mbox{$\frac{1}{2}$}}
\newcommand{\ket}{\rangle}

\def \invisible{\mbox{$\rule{0mm}{1mm}$}}
\def \mathbox(#1){\invisible\ifmmode{{#1}}\else{\mbox{${#1}$}}\fi}
\def \mbf(#1){\mbox{\boldmath{$#1$}}}

\def\mbfp{\mbf(p)}
\def\mbfr{\mbf(r)}
\def\mbfR{\mbf(R)}

\def\rmN{{\rm N}}

\def\NDmassdiff{\mbox{$\delta M_{{\rm N}\Delta}$}}
\def\SLJ(#1,#2,#3){\mathbox({^{#1}\!#2_{#3}})}

\def\VIII{\mathbox(V_{\rm III})}

\def\FRAC#1#2{\leavevmode\kern-.em
\raise.5ex\hbox{\the\scriptfont0 #1}\kern-.em
/\kern-.15em\lower.25ex\hbox{\the\scriptfont0 #2}}
\newif\ifrefphysrev

\def\refNP{\refphysrevfalse
           \typeout{** Reference: Nucl Phys format}}
\refphysrevtrue

\def \vol(#1,#2,#3){\ifrefphysrev{{\bf {#1}}, 
{#3} ({#2})}\else{{{\bf {#1}}({#2}){#3}}}\fi}
\def \NP(#1,#2,#3){Nucl.\ Phys.\          \vol(#1,#2,#3)}
\def \PL(#1,#2,#3){Phys.\ Lett.\          \vol(#1,#2,#3)}
\def \PRL(#1,#2,#3){Phys.\ Rev.\ Lett.\   \vol(#1,#2,#3)}
\def \PRp(#1,#2,#3){Phys.\ Rep.\          \vol(#1,#2,#3)}
\def \PR(#1,#2,#3){Phys.\ Rev.\           \vol(#1,#2,#3)}
\def \PTP(#1,#2,#3){Prog.\ Theor.\ Phys.\ \vol(#1,#2,#3)}
\def \ibid(#1,#2,#3){{\it ibid.}\         \vol(#1,#2,#3)}
\makeatletter
\def\scriptsize{\@setsize\scriptsize{14.5pt}\xipt\@xipt
\abovedisplayskip 11\p@ plus3\p@ minus6\p@
\belowdisplayskip \abovedisplayskip
\abovedisplayshortskip  \z@ plus3\p@
\belowdisplayshortskip  6.5\p@ plus3.5\p@ minus3\p@
\def\@listi{\leftmargin\leftmargini
\parsep 4.5\p@ plus2\p@ minus\p@ \itemsep \parsep
\topsep 9\p@ plus3\p@ minus5\p@}}
\def \@magscale#1{ scaled \magstep #1}
\def \half(#1){\mathbox(\frac{#1}{2})}
\def \ninej(#1,#2,#3,#4,#5,#6,#7,#8,#9){\mathbox(\left\{\matrix 
     {#1&#2&#3\cr#4&#5&#6\cr#7&#8&#9\cr}\right\})}
\newif\ifnoncomplete

\noncompletefalse

\def\@cite#1#2{\unskip\nobreak\relax
    {[#1]}} 

\def\citenum#1{{\def\@cite##1##2{##1}\cite{#1}}}
\def\citea#1{\@cite{#1}{}}
 
\newcount\@tempcntc
\def\@citex[#1]#2{\if@filesw\immediate\write\@auxout{%
\string\citation{#2}}\fi
  \@tempcnta\z@\@tempcntb\m@ne\def\@citea{}\@cite{\@for\@citeb:=#2\do
    {\@ifundefined
       {b@\@citeb}{\@citeo\@tempcntb\m@ne\@citea\def\@citea{,}%
{\bf ?}\@warning
       {Citation `\@citeb' on page \thepage \space undefined}}%
{\setbox\z@\hbox{\global\@tempcntc0\csname b@\@citeb\endcsname\relax}%
     \ifnum\@tempcntc=\z@ \@citeo\@tempcntb\m@ne
       \@citea\def\@citea{,}\hbox{\csname b@\@citeb\endcsname}%
     \else
      \advance\@tempcntb\@ne
      \ifnum\@tempcntb=\@tempcntc
      \else\advance\@tempcntb\m@ne\@citeo
      \@tempcnta\@tempcntc\@tempcntb\@tempcntc\fi\fi}}\@citeo}{#1}}
\def\@citeo{\ifnum\@tempcnta>\@tempcntb\else\@citea\def\@citea{,}%
  \ifnum\@tempcnta=\@tempcntb\the\@tempcnta\else
   {\advance\@tempcnta\@ne\ifnum\@tempcnta=\@tempcntb %
\else \def\@citea{--}\fi
    \advance\@tempcnta\m@ne\the\@tempcnta\@citea\the\@tempcntb}\fi\fi}

\def\affiliation#1{\gdef\@affiliation{#1}}
 
\def\and{\cr \makebox[0in]{\rule[-1cm]{0mm}{1cm}and } \cr}

\def\maketitle{\par
 \begingroup
 \def\thefootnote{\fnsymbol{footnote}}
 \def\@makefnmark{\hbox
 to 0pt{$^{\@thefnmark}$\hss}}
 \if@twocolumn
 \twocolumn[\@maketitle]
 \else \newpage
 \global\@topnum\z@ \@maketitle \fi\thispagestyle{plain}\@thanks
 \endgroup
 \setcounter{footnote}{0}
 \let\maketitle\relax
 \let\@maketitle\relax
 \gdef\@thanks{}\gdef\@author{}\gdef\@title{}
 \gdef\@affiliation{} \let\affiliation\relax	%
 \let\thanks\relax}

\def\@maketitle{\newpage
 \null
 \vskip 0em plus 2em minus 0em     
 \ifx\@date\@empty\else
   \begin{flushright}
    {\ifnoncomplete(\today)
     \else{{\normalsize \@date}\\}\fi}      
   \end{flushright}
   \vskip 3em plus 2em minus 2em   
 \fi
 \begin{center}
  {\Large \@title \par}     
  \vskip 3em plus 1em minus 1.5em  
  {
   \lineskip .5em plus 0em minus .3em   
   \begin{tabular}[t]{c}\@author\\
   \end{tabular}\par}
  \vskip 0.5em plus 1em minus 1.5em  
  { \sl \@affiliation \par}
\end{center}
 \par
 \vskip 6em plus 2em minus 4em}     
 
\def\abstract{\if@twocolumn
\section*{Abstract}
\else \normalsize
\fi}
 
\def\endabstract{\if@twocolumn\fi\par\clearpage}

 \makeatother
\refNP
   \input epsf           
\begin{document}
\title{
Effects of $\Sigma$N channels 
on the Spin-Orbit Force in $\Lambda$-hypernuclei
}
\author{Sachiko Takeuchi}
\affiliation{
Japan College of Social Work,
Kiyose, Tokyo 204-8555,
Japan}
\date{\today}
\maketitle
\baselineskip=20pt

\noindent
{\bf Abstract:~~}
Effects of the $\Sigma$N channels  
on the 
spin-orbit
force in $\Lambda$-hypernuclei
are investigated.
It is found that the
cancellation between the
symmetric and the antisymmetric $\Lambda$N spin-orbit force
becomes more complete
when the contribution from the
$\Sigma$N channels is included.
This may be the one of the reasons why
the observed LS splitting
of the $\Lambda$ single particle energy is very small.
\\

\noindent
PACS: 
12.39.Jh, 
13.75.Ev, 
21.30.Fe  
\\
Keywords:
Spin-orbit force,  
hyperon-nucleon interactions, Lambda-hypernuclei, quark model
\\

\newpage

The spin-orbit (LS) force in the two-baryon interaction has been investigated 
extensively.  
Information on the two-nucleon (NN) interaction can be directly obtained from
the NN scattering observables. 
As is well-known,
all the available phase shift analyses have concluded that there is 
a strong 
spin-orbit force between nucleons \cite{SAID,NijmPSA}.
The LS force in the hyperon-nucleon (YN) interaction, however, is not 
well-investigated
by the direct scattering experiments.
From the observed single particle energy levels of $\Lambda$-hypernuclei,
it is considered that the spin-orbit force 
between $\Lambda$ and the nucleon is very small 
comparing to that between two nucleons \cite{exp1,E930E419,theoM}.

The spin-orbit force of the YN interaction can be divided into 
the symmetric and the antisymmetric ones (SLS and ALS):
\begin{equation}
\calO_{{SLS \atop ALS}}
= {\sigma_i \pm \sigma_j \over  2} \cdot i [\mbfr_{ij} \times \mbfp_{ij}] ~,
\end{equation}
where 
$\mbfr_{ij}=\mbfr_i-\mbfr_j$ is the relative coordinate of the baryons and 
$\mbfp_{ij}=(m_j\mbfp_i-m_i\mbfp_j)/(m_i+m_j)$.
ALS becomes important in systems with strangeness.
The main part of ALS comes from the (symmetric) spin-orbit force 
in the quark-quark interaction 
and from the F/D difference 
of the vector-meson exchange, both of which 
survive at the flavor SU(3) limit  \cite{Tani,TMTO00}.
The size of ALS can be comparable to the
size of SLS
though it depends strongly on the channels. 

From a quark-model viewpoint,
the size of SLS between 
$\Lambda$ and the nucleon
can be comparable to that between two nucleons.
The spin-orbit force of the $\Lambda$ particle, however,
becomes small due to the SLS and ALS cancellation.
This is because
the s-quark alone among the quarks in the $\Lambda$ particle
produces the $\Lambda$N spin-orbit 
force:
the combined spin of the u- and d-quark in the $\Lambda$ particle
is zero.  
This small LS is divided artificially into SLS and ALS,
which cancel each other \cite{TMTO00,Ta98}.

The ALS and SLS cancellation in the $\Lambda$N channels, however, 
may not be enough to explain the observed small LS splitting of
 the $\Lambda$ single particle energy in hypernuclei.
In ref.\ \cite{RSY99}
several reasons why the LS splitting is small are discussed.
In this letter, we show that the
coupling to the $\Sigma$N channels
also affects the LS splitting of 
$\Lambda$-hypernuclei largely;
effective LS arising from the coupling to the $\Sigma$N channels
reduces the 
splitting of
 the $\Lambda$ single particle energy 
so that it becomes very small.

The model we employ here to investigate the above matter is the quark cluster model
(QCM).
This model with the instanton effects can 
explain the spin-orbit nature 
both in the two-baryon systems and in the single baryon mass spectrum
from the same quark hamiltonian,
and therefore enables us to investigate the spin-orbit force from 
more fundamental viewpoints
\cite{TMTO00,Ta98,Ta94}.

The present quark model  
contains four terms in the hamiltonian: 
the kinetic term, $K_{q}$, the confinement term, $V_{\rm conf}$, and the
one-gluon exchange (OGE) term, $V_{\rm OGE}$, and 
the instanton-induced interaction (\III) term $\VIII$, viz.
\cite{TMTO00,Ta94,OT91},
\begin{eqnarray}
H_{\rm quark} &=& K_{q}+(1-\piii)V_{\rm OGE} + \piii \VIII\ + V_{\rm conf} ~.
\label{eq1}
\end{eqnarray}
The parameter $\piii$ 
represents the relative strength of the
spin-spin part of \III\ to OGE.
It corresponds to the rate of the contribution from \III\ to  
the $S$-wave N-$\Delta$
mass difference, $\NDmassdiff$.

The YN interaction can be obtained from QCM as follows
\cite{TMTO00,Sh89,Ok95,PTP}.
The wave function is restricted as
\begin{eqnarray}
	\Psi_{6q} & = & \calA_{q} \{ \phi_{B} \phi_{B'} \chi(\mbfR) \} ~.
\label{quarkwf}
\end{eqnarray}
where $\calA_{q}$ is an antisymmetrizing operator for quarks, 
$\phi_{B}$ is the wave function of the baryon $B$,
and $\mbfR=(\mbfr_{1}+\mbfr_{2}+\mbfr_{3}-\mbfr_{4}-\mbfr_{5}-\mbfr_{6})/3$.
By integrating out the internal coordinates, we have the RGM equation,
\begin{eqnarray}
	(H - E N) \chi & = & 0 ~,
	\label{RGMeq}
\end{eqnarray}
where $H$ is the hamiltonian kernel and $N$ is the normalization kernel,
both of which are nonlocal.
The above equation can be rewritten as
\begin{eqnarray}
(\overline{H}-E) \overline{\chi}& = & 0
	\label{RGMeqBar} 
\end{eqnarray}
with
\begin{eqnarray}
	\overline{H} & = & N^{-1/2} H N^{-1/2}
	\label{Hbar}  \\
	\overline{\chi} & = & N^{1/2}\chi ~.
	\label{chibar} 
\end{eqnarray}
Thus, the short range part of the YN potential can be defined by
\begin{eqnarray}
	V_{{\rm QCM}} & = & \overline{H} - K_{0} ~,
	\label{VQCM}  
\end{eqnarray}
where $K_{0}$ is the kinetic term with $M_{B}=\sum m_{q}$.
The realistic YN hamiltonian can be obtained as
\begin{eqnarray}
	H_{{\rm YN}} &=& K + V_{{\rm QCM}}+V_{{\rm mesons}}~,
	\label{YN} 
\end{eqnarray}
where $K$ is the kinetic term with the observed baryon masses,
 and $V_{{\rm mesons}}$ is the one-boson exchange
potential with an appropriate form factor \cite{TMTO00}.
This hamiltonian can be used in usual Schr\"odinger equation:
\begin{eqnarray}
	(H_{{\rm YN}} - E) \psi &=& 0~,
	\label{YNeq}
\end{eqnarray}
with $\int |\psi|^{2} = 1$.

When there are one or more forbidden states due to the quark 
Pauli-blocking,
the normalization kernel does not have an inverse operator.
One has to remove the
forbidden states from the wave function $\psi$ in eq.\ (\ref{YNeq})
so that the operator $N^{-1/2}$ in eq.\ (\ref{Hbar}) 
becomes well-defined.
The YN ($I$=1/2 $S$=$-$1) $P$-wave channel in the present issue,
however, does not have such a state,
and is free from the above difficulty
when only the $\Lambda$N and $\Sigma$N channels are considered.
Eq.\ (\ref{YNeq}) is equivalent to eq.\ (\ref{RGMeq}) 
except for $V_{{\rm mesons}}$ and the choice of the baryon masses
in the kinetic term.

The $\Lambda$N interaction which includes the effects of the $\Sigma$N
channels  is obtained as:
\begin{eqnarray}
{\tilde H}_{\Lambda\rmN}(E_{0}) &=& H_{11} 
- H_{12} 
(H_{22} - E_{0})^{-1}
H_{21}  ~.
	\label{LNA} 
\end{eqnarray}
Here, $H_{ij}$ is the $(i,j)$-component of $H_{{\rm YN}}=\{H_{ij}\}$ in eq.\ 
(\ref{YN}), 
where the channel 1 [2] denotes the $\Lambda$N [$\Sigma$N] channel.
The solution of the equation,
$\{\tilde H_{\Lambda \rmN}(E_{0}) - E\}\psi=0$,
is the same as that of the coupled channel equation when $E_{0}$ 
equals to $E$;
the obtained scattering phase shifts are the same at 
$E=E_{0}$.
The hamiltonian of the single channel calculation, $H_{\Lambda 
\rmN}$, is
obtained by following the same procedure
from eq.\ (\ref{quarkwf}) to eq.\ (\ref{YN}) 
with the space restricted to the $\Lambda$N channels.
The difference, $\delta H \equiv \tilde H_{\Lambda\rmN} - H_{\Lambda\rmN}$,
is considered to be the correction 
due to the effects of the $\Sigma$N channels.

We use the parameter sets QCM-B, -C and -D, which were 
reported in ref.\ \cite{TMTO00}, to investigate the issue.
The QCM-B and -C contains no instanton effects ($\piii$=0). 
On the other hand,
$\piii$ is equal to 0.4  in QCM-D,
which is consistent to the value that 
gives the observed $\eta$-$\eta'$ mass 
difference and that gives the small LS splitting in the single baryon mass 
spectrum.
All the parameter sets 
contain the flavor singlet and octet, scalar- pseudoscalar- and vector-meson exchange
in $V_{{\rm mesons}}$.
Parameters in the quark part and in the meson part
are determined so that the model reproduces the
NN scattering data
and the cross section of the $\Lambda$p elastic scattering
in the 
low-energy region.
The parameter sets used in
the meson-exchange part of QCM-C and -D are the best fit each for
$\piii=0$ and for $\piii=0.4$, respectively.
The coupling constants of the vector mesons in QCM-B 
are kept to the same values as those in Nijmegen SC97f
though their strength is different from the original one due to the
difference of the cut-off method.

In fig.\ 1, 
we plot the phase shifts for the
$\Lambda$N \SLJ(3,P,J) channels given by QCM-D.
The solid lines correspond to the $\Lambda$N-$\Sigma$N coupled-channel calculation.
The phase shift of the \SLJ(3,P,1) channel, 
where the mixing of $\Lambda$N \SLJ(1,P,1) and $\Sigma$N \SLJ(3,P,1) 
channels by the ALS force becomes important,
oscillates largely at the 
$\Sigma$N threshold.
The dotted lines correspond to the single-channel calculation 
with $H_{\Lambda\rmN}$.
The dashed lines correspond to the results of single-channel 
calculation by 
$\tilde H_{\Lambda\rmN}$ with $E_{0}$ = 10 MeV;
 the effective
hamiltonian reproduces
the coupled-channel results well except for the threshold behavior.
\begin{figure}[tbp]
	\centering
	\includegraphics{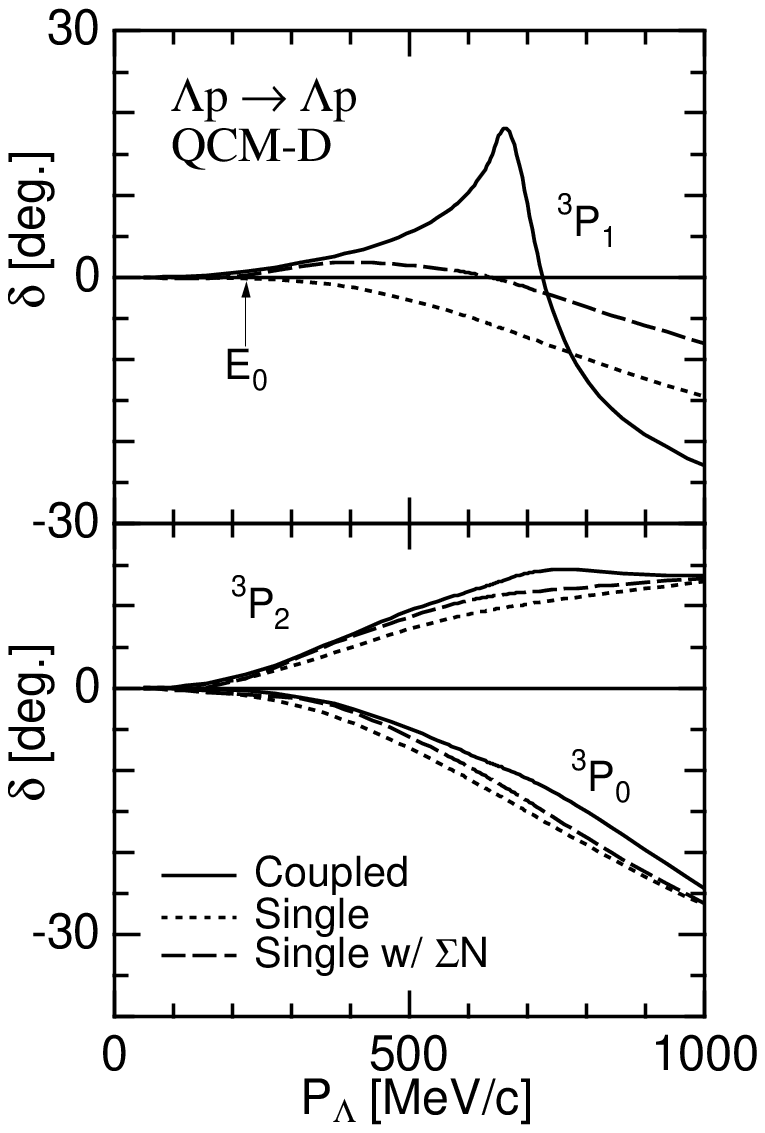}
	\caption[fig1]{Phase shifts of the $\Lambda$N $P$-wave channels.\\
The solid lines are for the $\Lambda$N-$\Sigma$N coupled-channel 
calculation, the dotted lines for the $\Lambda$N single-channel 
calculation, the dashed lines for the single-channel calculation 
with $\tilde H_{\Lambda\rmN}$($E_{0}$=10MeV) (see text).
}
	\label{fig1}
\end{figure}

To see rough size of the effects of the $\Sigma$N channels
on the $\Lambda$ hypernuclei,
we evaluate the spin-orbit part of the hamiltonian.
\begin{table}[tbp]
	\centering
	\caption{Evaluation of LS components}
\catcode`?=\active\def?{\phantom{$-$}}
\setlength{\tabcolsep}{1mm}
	\begin{tabular}{l|rrr|rrr|rrr}
		\hline
		\hline
		 & \multicolumn{3}{c|}{QCM-B} &\multicolumn{3}{|c|}{QCM-C} &  \multicolumn{3}{|c}{QCM-D}   \\
		 & (i) & (ii) \rule{5mm}{0cm}& (iii) & (i) & (ii) & (iii) &   (i) & (ii) & (iii) \\
		\hline
${H}_{\Lambda\rmN}{}_{SLS}$ & $-$1.37 & $-$1.03 ($-$0.56) & $-$0.50 & $-$1.62 & $-$1.22 & $-$0.59 & $-$1.48 & $-$1.20 & $-$0.58\\
${H}_{\Lambda\rmN}{}_{ALS}$ &    0.45 &    0.31 ?(0.07)   &    0.15 &    0.41 &    0.26 &    0.12 &    0.30 &    0.22 &    0.10\\
${H}_{\Lambda\rmN}{}_{LS}$  & $-$0.91 & $-$0.72 ($-$0.48) & $-$0.35 & $-$1.21 & $-$0.95 & $-$0.46 & $-$1.18 & $-$0.98 & $-$0.48\\
		\hline
$\delta H_{SLS}$            &    0.65 &    0.58 ?(0.50)   &    0.27 &    0.62 &    0.56 &    0.26 &    0.59 &    0.54 &    0.25\\
$\delta H_{ALS}$            &    0.88 &    0.67 ?(0.07)   &    0.27 &    0.67 &    0.50 &    0.20 &    0.58 &    0.45 &    0.18\\
\hline
\rule{0cm}{1.2em}$\tilde H_{\Lambda\rmN}{}_{LS}$ 
                            &  0.62 &    0.53 ?(0.10)   &    0.19 &    0.08 &    0.10 &    0.00 & $-$0.01 &    0.01 & $-$0.04\\
		\hline
		\hline
	\end{tabular}
	\label{table1}
\end{table}
The matrix elements of SLS  and ALS  
each for 
${H}_{\Lambda\rmN}$ and
$\delta H_{\Lambda\rmN}$
are listed in table 1
together with the total LS, $\tilde H_{\Lambda\rmN}{}_{LS}$.
The wave functions used for this estimate
are (i) the harmonic oscillator wave function 
with the size parameter of $b_{B}=$ 1.35 [fm],
(ii) the QCM wave function 
with the 
rms which corresponds to the one with $b_{B}=$ 
 1.35 fm, and (iii) that of 1.6 fm.
To obtain the wave functions of (ii) and (iii),
we solve the equation
with an extra gaussian potential with the central part of 
$H_{\Lambda\rmN}$,
\begin{eqnarray}
\{H_{\Lambda \rmN, cent}  +U_{0} R^{2}  - E\}\psi=0 ~.
\label{cent}
\end{eqnarray}
To see the effects of the 
difference in the hamiltonian clearly,
the orbital part of the obtained wave function for the \SLJ(3,P,{}) channel
is used as both of that for the \SLJ(3,P,J) and for the \SLJ(1,P,1) channels.
The value of $U_{0}$ is taken so that $\bra R^{2}\ket=5b_{B}^{2}$ with 
 $b_{B}=$ 
 1.35 fm or  1.6 fm.
This term also appears in the denominator of the second term of $\tilde H_{\Lambda\rmN}$.
By removing the term $U_{0} R^{2}$ from the denominator,
the correction listed in the table becomes larger by about 10\% $\sim$ 20\%.
The eigen energy of eq.\ (\ref{cent}) is used for $E_{0}$ in 
$\tilde H_{\Lambda\rmN}(E_{0})$; when the eigenvalue of each 
channel is used as $E_{0}$,
the results change by less than 1\% 
in the present case.

It is found that 
the introduction of the $\Sigma$N channels reduces SLS and enhances ALS.
As a result, the total LS, $\tilde H_{\Lambda\rmN}{}_{LS}$,
which corresponds to the $\Lambda$ LS splitting, 
becomes very small both in QCM-C and -D.
Let us emphasize that the small $\tilde H_{\Lambda\rmN}{}_{LS}$ is 
preferred by the experiments.

It is interesting to see whether 
this mechanism of cancellation due to the $\Sigma$N channels only 
occurs
in a quark model.
We have not investigated the issue
by using the meson-exchange models.
For the reference, however, 
the results with no quark spin-orbit force are listed
in parentheses for the QCM-B (ii) case in table 1.
Though the absolute value is different, especially for ALS, which is 
weakened by the cut off,
the fact that the effects of the $\Sigma$N channels reduce
the total LS seems valid for the meson LS.

The correction can be written as
\begin{eqnarray}
	\delta H & = & (H_{11}-H_{\Lambda\rmN}) + H_{12}(H_{22}-E_{0})^{-1}H_{21} ~.
	\label{delH}
\end{eqnarray}
The first term of rhs
is the contribution of the $\Sigma$N channels through
the normalization kernel,
which does not appear in the meson-exchange models.
Its contribution, however,
seems minor: it enhances SLS only by less than 5\%.
This term for ALS is $-$0.09
for the case (ii) of QCM-D, for example;
it is reverse
 and still considerably
 smaller comparing to the second term.

The second term of eq.\ (\ref{delH}), which is common to the meson-exchange models,
 gives a major contribution to the correction, $\delta H$.
This contribution to ALS can be divided into four terms.
Namely, each pair of $(S,S')$ in 
\begin{eqnarray}
\sum_{S,S'}
\bra \Lambda\rmN \SLJ(3,P,1) | H_{12}|S\ket
\bra S | (H_{22}-E_{0})^{-1}| S' \ket
\bra S'| H_{21}| \Lambda\rmN \SLJ(1,P,1) \ket  ~.
\label{HZH}
\end{eqnarray}
where $|S\ket$ stands for $| \Sigma\rmN \SLJ(2S+1,P,1) \ket$.
It is found that all the four terms contribute additively to $\delta 
H_{ALS}$.
The channel dependence of ALS is determined only by the flavor SU(3)
symmetry and does not depend on the origins of ALS \cite{Tani}.
That is, the strength of ALS in a certain channel is essentially governed
by the symmetry and one dynamical matrix element.
One the other hand,
the factor $\bra S | (H_{22}-E_{0})^{-1}| S' \ket$ should not change 
much 
if one uses a realistic YN potential.
Therefore,
ALS of the system  should  enhance similarly  by eq.\ (\ref{HZH}),
provided that the $\Lambda$N-$\Sigma$N off-diagonal matrix element
given by QCM is similar to the one with the meson-exchange potentials;
which is likely because the models maintain the flavor SU(3)
symmetry approximately.

The situation for SLS is vague, because
the channel dependence of SLS is not determined only by the symmetry
unlike that of ALS.
The sign of meson SLS, however, is the same as that of SLS
originated from OGE or III interacting between quarks
except for the $\Sigma$N-$\Sigma$N ($I=$1/2) \SLJ(3,P,J) channel \cite{Tani}.
Therefore, we expect that the similar correction arises
also for SLS.

As is seen in table 1, the size of the effects may vary.
Here, at least for the best fit models which reproduces the NN scattering data
and the $\Lambda$N low energy data, it seems the LS becomes almost zero
by introducing the present effects.
Recently, a work investigating $\Lambda$ LS in hypernuclei
by using the $G$-matrix approach
with a quark cluster model has been reported \cite{KFFNS99};
they have concluded that the small LS splitting for $\Lambda$ in hypernuclei
can be obtained from their model.
The similar mechanism to the present work probably occurs in their
approach.

In this letter we investigate the effects of the coupling
to the $\Sigma$N channels on the $\Lambda$N spin-orbit force
employing the quark cluster model.
It is found that the coupling reduces the LS force
considerably.
This mechanism probably contributes to
the observed small LS splitting
of the $\Lambda$ single particle energy.
\bigskip

The author would like to thank Dr.\ K.\ Shimizu for valuable
discussions.
This work is supported in part by
the Grant-in-Aid for Scientific Research (C)(2)11640258
and (C)(2)12640290
of Ministry of Education, Science, Sports and Culture of Japan.
%

\end{document}